\documentclass[eqsecnum,preprint,prd]{revtex4}

\usepackage{amssymb}

\usepackage{graphicx}

\begin{document}

\title{On exact solutions for  quintessential (inflationary) cosmological models with exponential potentials }

\author{Ester Piedipalumbo and Paolo Scudellaro}
\email[E-mail at: ]{ester@na.infn.it}
\affiliation{Dipartimento di Scienze Fisiche, Universit\`{a} Federico II
and  Istituto Nazionale di Fisica Nucleare, Sez. di Napoli,\\
Complesso Universitario di Monte S. Angelo,\\ Via Cintia, Ed. 6, I-80126 Napoli, Italy}

\author{Giampiero Esposito and Claudio Rubano}
\affiliation{Istituto Nazionale di Fisica Nucleare, Sez. di Napoli,\\
Complesso Universitario di Monte S. Angelo,\\ Via Cintia, Ed. 6, I-80126 Napoli, Italy}

\vspace{0.4cm}
\date{\today}

\begin{abstract}
We first study dark energy models with a minimally--coupled scalar field and exponential potentials, admitting
exact solutions for the cosmological equations: actually, it turns out that for this class of potentials the
Einstein field equation exhibit alternative Lagrangians, and are completely integrable and separable
(i.e. it is possible to integrate the system analytically, at least by quadratures). We analyze such solutions,
especially discussing when they are compatible with a late time quintessential expansion of the universe. As a
further issue, we discuss how such quintessential scalar fields can be connected to the inflationary phase,
building up, for this class of potentials, a quintessential inflationary scenario: actually, it turns out that
the transition from inflation toward late--time exponential quintessential tail admits a kination period, which
is an indispensable ingredient of this kind of theoretical models. All such considerations have also been done
including radiation into the model.

\end{abstract}

\pacs{98.80.Cq, 98.80.Hw, 04.20.Jb}

\maketitle
\bigskip
\vspace{2cm}

\section{Introduction}

Scalar fields have been used extensively in cosmology since inflationary theories were first
conceived in 1981 \cite{guth}. They are in fact useful to fill in the universe with a negative--pressure
content, hence giving the typical inflationary behavior, with its exponentially increasing scale factor $a(t) \sim
\exp(H_0t)$, where $H_0$ is the present value of the Hubble parameter. Scalar fields have been used both in
minimal and non minimal coupling with geometry. (An overview of scalar--tensor theories is given in Refs.
\cite{dam} \cite{de1}; a more updated review can be found in Ref. \cite{STreview}.)

When observational results on Type Ia supernovae (SNeIa) have strongly supported the possibility that the
universe is now in an accelerated stage of expansion \cite{per1} \cite{per2} \cite{rie} \cite{gar}, it has
become necessary to consider again a negative--pressure component. The late-time inflation so generated is
however {\em soft} with respect to the earlier one and is again based on the dominance of at least one scalar
field referred to as {\em quintessence} \cite{ost} \cite{cal} \cite{zla} \cite{ste1} \cite{wan} or, earlier, as
{\em x--field} \cite{tur1} \cite{chi}. This gives rise to a sort of energy component which only appears
measurable through gravitational effects. There is in fact a wide agreement that today a form of such a {\em
dark energy} \cite{van} \cite{tur2} has to be taken into account in any realistic cosmological model. Seen as a
constant $\Lambda$--term \cite{car1} \cite{car2} \cite{sah1}, dark energy density has been usually seen as the
vacuum energy density \cite{rob}, leading therefore to relevant problems due to huge discrepancies between theory and
experiments \cite{wei} \cite{car2} \cite{sah1} \cite{rugh}.

In this context, the scalar field in itself has gained new attention, and speculations about its nature and
evolution have got new strength. Papers like Refs. \cite{rat} and \cite{pee} have been reconsidered again under
a new light, as they show much of the role a scalar field could play when one wants to have an accelerated
expansion today. Many models have thereof been constructed, first of all trying to consider more appealing kinds
of potentials driving the dynamics of the scalar field. Among others, the potential has in fact to be seen as an
important ingredient of all the involved theoretical models, and many are the trials to give prescriptions in
order to reconstruct its form according to observational data (see Refs. \cite{sai} \cite{rub1}, for instance;
but, for more theoretical considerations, see also Refs. \cite{lid} \cite{cal} \cite{zla} \cite{ste1} \cite{bra}
\cite{bin}). Here we take into account a specific class of potentials useful for the present evolution of the
universe, later investigating all the relative involved cases so derived in the cosmological equations. The
exponential--type potential has received much attention since the late 80's \cite{bar} \cite{bur} \cite{rat},
(but see also Refs. \cite{lid} \cite{wet1} \cite{wet2} \cite{cop} \cite{fer} \cite{fab} \cite{barr} \cite{sah2}
\cite{bra} \cite{bat}, \cite{ack}, just to cite only some other old papers). In Refs. \cite{rub2} \cite{rub3} we have found
general exact solutions for two classes of exponential potentials for a scalar field minimally coupled to
gravity, in presence of another dust component (ordinary non--relativistic pressureless baryonic matter), also
showing \cite{pav} that such solutions seem to fit SNeIa data given in Ref. \cite{per2}. (See Ref. \cite{dem}
for much more on confrontation of theoretical predictions with observational data.) Furthermore, in Ref.
\cite{keko} other approximate and exact solutions for cosmology with exponential potentials can be found, while
up to more recently it is still widely referred to (as, for example, in Ref. \cite{batsapa}, where the Noether
Symmetry Approach \cite{de2} \cite{de1} is used to probe the nature of dark energy, well underlining the role of
the exponential potential in such a task). This seems to us a good motivation to investigate also on
generalizations of this kind of potentials. (As a matter of fact, it is often studied also the case with the
hyperbolic sine potential, which can easily be assimilated with some situations studied below).

Moreover, also for such a class of generalized exponential potentials it is possible to exhibit some
\textit{special} properties: it turns out, actually, that the Einstein field equations are integrable (in
the Liouville sense) and separable, that is they can be analytically integrated, at least by quadrature. Finally, it is worth noting
that our class of exponential potentials can be selected by finding the most general variables transformation
which diagonalizes the scalar field kinetic-energy form, leaving the transformed Lagrangian \emph{simple}.

Therefore, among other considerations, we then generalize the exponential potential to the extent allowed by the
particular technique used in Refs. \cite{rub2} \cite{rub3} in order to integrate cosmological equations, still
bearing in mind that, in order to describe the present universe, solutions should always allow a late--time
acceleration. Such a technique is generally known as the Noether Symmetry Approach to cosmology \cite{de2},
\cite{de1}. (Note that in Refs. \cite{de2}, \cite{de1}, such an application of the Noether Symmetry Approach
already leads to prefer the exponential--type potentials.) As in Refs. \cite{rub2} and \cite{rub3} we anyway do not
completely adopt it here, limiting ourselves to import from it only the change of variables needed to solve
equations. As a matter of fact, we have to point out that this transformation (together with the choice of a
specific kind of potential) has to be considered as the main drawback of the approach above, since it leads to a
point symmetry which probably applies only to the background evolution, having nothing to do with the actual
symmetries of the full theory. We analyze the exact solutions we find in this way, discussing when they are compatible
with a late time quintessential expansion of the universe.

As a further step, we also consider and illustrate a possible scenario in the framework of the
\textit{quintessential inflation paradigm} with scalar fields, where  an inflationary  potential drives also the quintessential phase of the scalar field evolution, by means of an exponential form of its late time tail.
In this connection, we discuss how such an evolution mechanism for the scalar field potential can be compatible
with the so-called inflation--kination transition, when the field energy density is dominated by the kinetic
energy of the scalar field $\varphi$. All this clearly leaves apart the greater complexity involved by the more general
forms of the potential studied above, and will be the subject of a forthcoming paper trying to discuss more in general
the quintessential inflation. Even if in this second paper we should be able to overcome the dichotomy present in this work,
where the two different parts of the paper could well live apart, we believe that it is nonetheless already interesting
to offer both the issues in a single paper, trying to unify the whole subject a little forcibly.

In Sec. II we introduce the class of potentials and the general cosmological setting for further considerations.
In Sec. III, we systematically derive, when possible, general exact solutions case by case. In Sec. IV we
discuss connections between exponential potentials quintessence and inflation. Finally, Sec. V is devoted to a
conclusive discussion.

\section{Cosmological models with (generalized) exponential potentials}

Motivated by the considerations above, let us continue to assume a
Friedmann--Lema${\hat{i}}$tre--Robertson--Walker (FLRW) metric and fix the curvature scalar $k = 0$, since the
spatially flat situation appears to be the most appropriate one according to the CMBR observational data (see
Refs. \cite{newCMB1} \cite{newCMB2} for recent information). In what follows we have to consider the period of
life of the universe after the decoupling time, in order to be realistic when taking the two components like
dust and scalar field $\varphi$ into account, like in Refs. \cite{rub2} \cite{rub3}.

If $\varphi$ is minimally coupled to gravity, and considering no special choice for constants, the cosmological
equations are written
\begin{equation}
3H^{2} = \frac{8\pi G}{c^{2}}(\rho _{m} + \rho _{\varphi})\,, \label{eq1}
\end{equation}
\begin{equation}
\dot{H} + H^{2} = -\frac{4\pi G}{c^{2}}(\rho _{m} + \rho _{\varphi} + 3(p_{m} + p_{\varphi}))\,, \label{eq2}
\end{equation}
\begin{equation}
\ddot{\varphi} + 3H\dot{\varphi} + V^{\prime}(\varphi) = 0\,. \label{eq3}
\end{equation}
The fluids filling the universe have equations of state given by
\begin{equation}
p_m = 0\,,\,\,\,\,\,\,\, p_{\varphi} = w_{\varphi}\rho_{\varphi}\,, \label{eq4}
\end{equation}
being $\rho_m = D a^{-3}$, where the parameter $D \equiv \rho_{m0}{a_0}^3$ is determined by the current
values of $\rho_m$ and $a$. Let us also recall that
\begin{equation}
\rho_{\varphi} \equiv \frac{1}{2}{\dot{\varphi}}^2 + V(\varphi)\,,\,\,\,\,\,\,\, p_{\varphi} \equiv
\frac{1}{2}{\dot{\varphi}}^2 - V(\varphi)\,, \label{eq5}
\end{equation}
and
\begin{equation}
w_{\varphi} \equiv \frac{{\dot{\varphi}}^2 - 2V(\varphi)}{{\dot{\varphi}}^2 + 2V(\varphi)}\,. \label{eq6}
\end{equation}

The cosmological equations can then be rewritten as
\begin{equation}
\left( \frac{\dot{a}}{a} \right)^{2} = \frac{8\pi G}{c^{2}}\left( Da^{-3} + \frac{1}{2}{\dot{\varphi}}^2 +
V(\varphi) \right)\,, \label{eq7}
\end{equation}
\begin{equation}
2\frac{\ddot{a}}{a} + \left( \frac{\dot{a}}{a} \right)^{2} = -\frac{8\pi G}{c^{2}}\left(
\frac{1}{2}{\dot{\varphi}}^2 - V(\varphi) \right)\,, \label{eq8}
\end{equation}
\begin{equation}
\ddot{\varphi} + 3\left( \frac{\dot{a}}{a} \right)\dot{\varphi} + V^{\prime}(\varphi) = 0\,. \label{eq9}
\end{equation}
It can be shown \cite{de2},\cite{de1} that the last two equations can also be deduced from an action principle
based on the {\em point} Lagrangian
\begin{equation}
{\cal L} = 3a{\dot{a}}^2 - \frac{8 \pi G}{c^2}\left[ a^3\left(\frac{1}{2} {\dot{\varphi}}^2 - V(\varphi) \right)
- D \right]\,. \label{eq10}
\end{equation}
In this way, cosmological dynamics can be considered on a space with two {\em coordinates} $a$
and $\varphi$, $\dot{a}$ and $\dot{\varphi}$ being the {\em velocities} \cite{de2} \cite{de1}. The fact that
${\cal L}$ has a constant additive term is understood by considering Eq. (\ref{eq7}), which can be seen as $E_{\cal
L} = 0$, where $E_{\cal L} \equiv {\frac{\partial {\cal L}}{\partial \dot{a}} + \frac{\partial {\cal L}}
{\partial \dot{\varphi}} - {\cal L}}$ is the so called {\em energy function} associated with ${\cal L}$. We find
the constant to be zero for physical reasons (i.e., induced by the homogeneous and isotropic limit of Einstein's
general field equations). In this way, when Eqs. (\ref{eq8}) and (\ref{eq9}) are solved, Eq. (\ref{eq7}) is
nothing but a constraint on the integration constants involved.

On defining
\begin{equation}
T \equiv 3a{\dot{a}}^2 - \frac{4 \pi G}{c^2}a^3 {\dot{\varphi}}^2\,,\,\,\,\,\, U \equiv - \frac{8 \pi
G}{c^2}(a^3 V(\varphi) + D)\,,
\end{equation}
Eq. (\ref{eq10}) formally becomes ${\cal L} = T - U$. This has been already noted in Ref. \cite{de2bis}, where
we found the most general transformation
\begin{equation}
a = f(z,w)\,\,\,\,\,\, \varphi = g(z,w)\,
\end{equation}
leading to a \emph{diagonalized} and simpler form of the \emph{kinetic energy}
\begin{equation}
T' = {\alpha}^2 {\dot{z}}^2 - {\beta}^2 {\dot{w}}^2\,,
\end{equation}
with $\alpha$ and $\beta$ nonnegative real numbers. This simplification in $T$ implies, of course, a
complication in the transformed $U'$. However in Ref. \cite{de2} we also found that there is at least one class
of potentials not giving such a complicated expression for $U'$, rendering, on the contrary, the transformed
${\cal L}'$ simpler than ${\cal L}$. It turns out that such potentials are exponentials:
\begin{equation}
V(\varphi) = V_0\left(A^2 \exp{(2C\varphi)} + B^2 \exp{(-2C\varphi)} - 2AB\right)\,,
\end{equation}
with $C \equiv (3\pi G)/c^2$ and $A$, $B$ real parameters. They in fact allow an exact integration of the
cosmological equations \cite{de2} \cite{de1}. Moreover, it is worth noting that, for such a class of exponential
potentials,
\begin{equation}
V(\varphi) = V_1 \exp{(2C\varphi)} + V_2 \exp{(-2C\varphi)} + \lambda_0\,,
\end{equation}
being $V_1$, $V_2$, $\lambda_0$  all free parameters, the Einstein field equations admit alternative
Lagrangians, which is a circumstance somehow exceptional for a dynamical system, with many meaningful
consequences: actually, it turns out that it is integrable (in the Liouville sense) and it is separable, i.e.
there is a suitable change of variables by which it is splitted into separated one--dimensional systems (which
are integrable by quadratures \cite{epl95}).

We thus feel motivated to continue considering exponential--like potentials, taking now the class of potentials
\begin{equation}
V(\varphi) = A^2 \exp{(\sigma \varphi)} + \epsilon B^2 \exp{(-\sigma \varphi)} + \lambda\,, \label{eq11}
\end{equation}
where $\epsilon = {\pm} 1$, $\sigma \equiv \sqrt{12 \pi G/c^2}$ is a fixed constant, $A^2$ and $B^2$ are
arbitrary nonzero parameters, and $\lambda \lesseqgtr 0$. Eq. (\ref{eq11}) generalizes the exponential
potentials already considered in Ref. \cite{rub2}, which are
\begin{equation}
V(\varphi) = B^2 \exp{(-\sigma \varphi)} \,\,\,\,\,, V(\varphi) = A^2 \exp{(\sigma \varphi)} + B^2 \exp{(-\sigma
\varphi)} \,. \label{eq11bis}
\end{equation}
These cases (where $\epsilon = + 1$ and $\lambda = 0$) are therefore omitted in this paper. Moreover, as noted,
the case with $\epsilon = + 1$ and $\lambda = -2AB$ has been already treated and discussed in Refs. \cite{de2}
\cite{de2bis} \cite{de1}, and it will not be touched upon again here.

Since the theory is invariant under $\varphi \rightarrow - \varphi$, the class of potentials in Eq. (\ref{eq11})
is the most general one of this type and also includes the first one shown in Eq. (\ref{eq11bis}). $V(\varphi) =
B^2 \exp{(-\sigma \varphi)}$ is in fact derived from the expression in Eq. (\ref{eq11}) just setting (besides
$\epsilon = +1$ and $\lambda = 0$) also $A^2 = 0$, or $B^2 = 0$ and $\varphi \rightarrow - \varphi$. (Of course,
in order to get exact general solutions, the treatment is different from the beginning for each one of the
potentials in Eqs. (\ref{eq11}) and (\ref{eq11bis}).)

As (among others) in Ref. \cite{rub2}, then, let us use again the transformation
\begin{equation}
a^{3}=\frac{u^{2} - v^{2}}{4}\,,\,\,\,\,\,\,\, \varphi = \frac{1}{\sigma }\log\left| \frac{B(u+v)}{A(u-v)}
\right|\,,  \label{eq12}
\end{equation}
leading to the new variables $u$ and $v$. Such a change of variables is invertible, provided that $a \neq 0$. Of
course, when the solution for the scale factor eventually has a zero in the future, we cannot follow it beyond
that point. We must in fact consider such a solution as good just \emph{between} this zero and the eventual
other one it can have in the past. (Later on, we will comment on this again.) Since we also want $a > 0$, Eq.
(\ref{eq12}) imposes the obvious restriction $u^2
> v^2$, or $(u + v)(u - v)
> 0$, giving $u > v$ when both $u$ and $v$ are nonnegative. This soon implies that $A$ and $B$ must have the
same signs (i.e., $AB > 0$). In what follows, we choose to consider $A > 0, B > 0, u > v$, and $u > 0, v > 0$.
We also disregard the absolute value sign in the expression for $\varphi$, which does not really limit the class
of potentials we are studying.
By virtue of  Eq. (\ref{eq12}), the potential in Eq. (\ref{eq11}) becomes
\begin{equation}
V(u,v) = AB\frac{(u + v)^2 + \epsilon (u - v)^2}{u^2 - v^2} + \lambda\,, \label{eq13}
\end{equation}
and we define
\begin{equation}
V_{+}(u,v) \equiv 2AB\frac{u^2 + v^2}{u^2 - v^2} + \lambda\, \label{eq14}
\end{equation}
for $\epsilon = + 1$, and
\begin{equation}
V_{-}(u,v) \equiv 4AB\frac{uv}{u^2 - v^2} + \lambda\, \label{eq15}
\end{equation}
for $\epsilon = - 1$.

In the case when $A^2 = 0$, anyway, we have to use the change of variables
\begin{equation}
a^{3}=uv\,,\,\,\,\,\,\,\, \varphi =-\frac{1}{\sigma }\log \frac{u}{v}\,, \label{eq15bis}
\end{equation}
as already done in Ref. \cite{rub2} when $\lambda = 0$. This case will be treated in detail in Sec. III, taking
into account the possible values of $\lambda$.

Now, transforming the other terms in Eq. (\ref{eq10}) by means of Eq. (\ref{eq12}), we can eventually arrive at
two expressions for the point Lagrangian, according again to the two opposite values of $\epsilon$,
\begin{equation}
{\cal L}_{+} \equiv {\dot{u}}^2 - {\dot{v}}^2 + \frac{{\sigma}^2}{2}\left[ (2AB + \lambda)u^2 + (2AB -
\lambda)v^2 + 4D \right]\,, \label{eq16}
\end{equation}
\begin{equation}
{\cal L}_{-} \equiv {\dot{u}}^2 - {\dot{v}}^2 + \frac{{\sigma}^2}{2}\left[ 4ABuv + \lambda(u^2 - v^2) + 4D
\right]\,\label{eq17}
\end{equation}
where the possibility of multiplying the Lagrangian by a costant, equal to $3$ in our case, has been exploited
In order to find equations and solutions for the cosmological model, at this point, it is better to consider the
various different cases separately, and this is what we are going to do in the next Section.

In the following, we fix four conditions (see Ref. \cite{rub3} for further details). First of all, we set the
origin of time by choosing $a(0)= 0$, paying attention to the fact that, even if this can be done generally, it is
not a real initial condition for $a(t)$. The initial time $t_in$ chosen for any specific model is in fact
arbitrary, to some extent, and mathematical solutions can also present more than one zero, eventually infinitely many. In such cases, one must be careful in choosing only one physically meaningful period of time for the
evolution of the universe, noting that we can accept no more than one zero at the beginning of time. An
eventual second zero occurring in the future has then to be recognized as a time of recollapse of the universe.
What happens later thus loses any meaning. We have thus to set $u(0) = 0$ \textit{or} $v (0) = 0$ in evaluating
initially Eq. (\ref{eq15bis}), being it arbitrary for the moment to decide which one is better on solid ground.
In Ref. \cite{rub2}, for instance, we decided to set to zero \textit{both}, so selecting the tracking solution.

The second condition that we assume is that the present time (age of the universe) is the unit of time,
$t_0 = 1$. Being $t_{in}$ unknown, this is not exactly the age of the universe, but the difference can be
considered irrelevant for our purposes; anyway, even if it were be possible to avoid this condition, we thus
get rid of a badly known quantity. Our third condition is to set $a_0\equiv a(t_0)= a(1) = 1$, which fixes the
normalization of $a$ as standard, while the last condition is to set $H(t_0 = 1)\equiv {\cal H}_0$. After the
choice of $t_0$, this latter parameter turns out to be of order $1$, even if it is not the same as the usual $h$.
As a matter of fact, since we are using an arbitrary unit of time, such a parameter does not give any
information on the observed value for $H_0$ \cite{rub3}.

As to $\lambda = 0$, as already said, some considerations are given in Ref. \cite{rub2}. When $\lambda =
A^2 = 0$ and $\epsilon = +1$, for instance, Eq. (\ref{eq15bis}) easily leads to general exact solutions for
$a(t)$ and $\varphi(t)$ \cite{rub2}. For $\lambda = B^2 = 0$, everything works the same, since there is symmetry
in the potential in Eq. (\ref{eq11}) with respect to a change of sign in $\varphi$. The situation where $A^2
\neq 0$, $B^2 \neq 0$, and $\epsilon = +1$ has also been partially examined in Ref. \cite{rub2}.

In the next Section, we will thus focus our attention mainly on the case with $A^2 \neq 0$, $B^2 \neq 0$ and
$\epsilon = - 1$, choosing not to consider at all the situation with $\lambda = A^2 = 0$ and $\epsilon = -1$,
since it involves an always negative potential and may also give a negative energy density, by virtue of  Eq.
(\ref{eq5}). Most of all, let us note that it never allows the pressure to be negative, which forbids the
possibility of describing the accelerated expansion we observe today.

We will also rule out the trivial case given by $V \equiv \lambda$, while the remaining situations will be dealt
with systematically.

\section{Solutions}

Once we have fixed our Lagrangian expressions in Eqs. (\ref{eq16}) and (\ref{eq17}), we can soon derive the
related equations for cosmology as Euler-Lagrange equations. If our transformation in Eq. (\ref{eq12}) then
works, such equations should turn out to be solvable, even if, of course, exact integration is not always easy. The
analysis strongly depends on the relative values assumed by the constants involved. Thus, in what follows we
separately discuss each situation generated, as a first step, by a different choice of $\lambda$ values. Then,
we pass to further investigate the features of the equations and their solutions, starting from consideration of
the two allowed values of $\epsilon$.

\subsection{The $\lambda = 0$ case}

As mentioned, this situation has already been partially treated in Ref. \cite{rub2}, where we consider only
three cases: i) $A^2 = 0$ and $\epsilon = +1$, ii) $B^2 = 0$, and iii) $A^2 \neq 0$, $B^2 \neq 0$, and $\epsilon
= +1$.
The case with $A^2 = 0$ and $\epsilon = -1$ has already been touched upon in Sec. II, noting
that it is not so interesting here in our considerations on present cosmology. On the other hand, when both
$A^2$ and $B^2$ vanish, the potential vanishes and $\varphi$ is free, giving $p_{\varphi} \equiv
\rho_{\varphi} \equiv {\dot{\varphi}}^2/2 > 0$, and $w_{\varphi} = 1$. This means that the scalar field behaves
like stiff matter and introduces a term $\propto a^{-6}$ in the cosmological equations, thus becoming soon
negligible in the expanding evolution of the universe. In this way, $\varphi$ would never produce a late-time
inflation, for instance, as instead recent observations seem to require.

There only remains one case to discuss, i.e.,
\begin{equation}
A^2 \neq 0\,,\,\,\,\,\,\,\, B^2 \neq 0\,,\,\,\,\,\,\,\, \epsilon = - 1\,. \label{eq19}
\end{equation}
With such assumptions, setting $\lambda = 0$ in Eq. (\ref{eq17}) yields
\begin{equation}
{\cal L}_{-} \equiv {\dot{u}}^2 - {\dot{v}}^2 + \frac{24 \pi G}{c^2}(ABuv + D)\,, \label{eq20}
\end{equation}
so that the related Euler-Lagrange equations are
\begin{equation}
\ddot{u} = {\omega}^2v\,,\,\,\,\,\,\,\, \ddot{v} = - {\omega}^2u\,, \label{eq21}
\end{equation}
where
\begin{equation}
{\omega}^2 \equiv \frac{12 \pi GAB}{c^2} = {\sigma}^2 AB\,. \label{eq22}
\end{equation}

It can be shown that the solution reads as
\begin{eqnarray}
u(t) & = & \left[ \frac{c_3 + c_4}{\sqrt{2}\omega} \cosh(\frac{\omega t}{\sqrt{2}}) + c_2 \sinh(\frac{\omega
t}{\sqrt{2}})
\right]\sin(\frac{\omega t}{\sqrt{2}}) \nonumber \\
     &   & \qquad \qquad \qquad - \left[ \frac{c_3 - c_4}{\sqrt{2}\omega}
\sinh(\frac{\omega t}{\sqrt{2}}) + c_1 \cosh(\frac{\omega t}{\sqrt{2}}) \right]\cos(\frac{\omega
t}{\sqrt{2}})\,, \label{eq23}
\end{eqnarray}
\begin{eqnarray}
v(t) & = & \left[ - \frac{c_3 - c_4}{\sqrt{2}\omega} \cosh(\frac{\omega t}{\sqrt{2}}) - c_1 \sinh(\frac{\omega
t}{\sqrt{2}})
\right]\sin(\frac{\omega t}{\sqrt{2}}) \nonumber \\
     &   & \qquad \qquad \qquad + \left[ \frac{c_3 + c_4}{\sqrt{2}\omega}
\sinh(\frac{\omega t}{\sqrt{2}}) + c_2 \cosh(\frac{\omega t}{\sqrt{2}}) \right]\cos(\frac{\omega
t}{\sqrt{2}})\,, \label{eq24}
\end{eqnarray}
so that we get two functions of time $t$ depending on four arbitrary parameters ($u_1$, $u_2$, $t_1$, and
$t_2$). By virtue of the energy constraint $E_{\cal L_{-}} = 0$, which is written
\begin{equation}
{\dot{u}}^2 - {\dot{v}}^2 - 2{{\omega}^2}uv - \frac{24 \pi G}{c^2}D = 0\,, \label{eq25}
\end{equation}
 Eq. (\ref{eq22}) yields
\begin{equation}
D = AB\left( \frac{{c_3}^2 -{c_4}^2}{2{\omega}^2} - c_1 c_2 \right)\,. \label{eq26}
\end{equation}
If we define $k_3 \equiv (c_3 + c_4)/(\sqrt{2}\omega)$ and $k_4 \equiv (c_3 - c_4)/(\sqrt{2}\omega)$ we then get
$D = AB(k_3 k_4 - c_1 c_2)$. The initial scale factor at $t = 0$ is such that the volume $a^3(t)$ is at the {\em
beginning} $a^3(0) = ({c_1}^2 - {c_2}^2)/4$, in light of Eq. (\ref{eq12}). For sake of simplicity, we can fix the
origin of time in such a way as to get a vanishing scale factor (as in Ref. \cite{rub2}).
This, of course, is completely arbitrary, and we must always remember that the real beginning of time for the
model under examination is actually afterwards, i.e. , at an after-decoupling moment which can be arbitrarily
delayed for the kind of considerations we are making here. This also means, then, that we are setting $c_1 = c_2$
or, possibly, $c_1 = c_2 = 0$ (since we have put $a(0) = 0$). Anyway, both choices still allow $D$ to be nonvanishing and positive, which we certainly require. Of course, this would be gained even taking $c_2 =
0$ simply. But let us here put $c_1 = c_2 = 0$, so that we can write
\begin{equation}
D = ABk_3 k_4\,, \label{eq26bis}
\end{equation}
from which we find that the two constants $k_3$ and $k_4$ have the same sign, i.e., $k_3 k_4 > 0$, implying that
 ${c_3}^2 > {c_4}^2$.

Let us also note that, from Eqs. (\ref{eq23}) and (\ref{eq24}), we get  $u(0) = - c_1$ and $v(0) = c_2$. Thus,
independently of all considerations above, Eq. (\ref{eq12})$_2$ implies
\begin{equation}
\varphi(0) = \frac{1}{\sigma}\log \frac{B(u(0) + v(0))}{A(u(0) - v(0))} = \frac{1}{\sigma}\log \frac{B(- c_1 +
c_2)}{A(- c_1 - c_2)} \,, \label{eq26ter}
\end{equation}
and the choice $c_1 = c_2$ or, equivalently, $a(0) = 0$, soon gives $\varphi (0) = 0$.

By substituting Eqs. (\ref{eq23}) and (\ref{eq24}) into the expressions in Eq. (\ref{eq12}) for $a^3(t)$ and
$\varphi(t)$, after some algebra we find the cosmological solutions
\begin{eqnarray}
a(t) & = & \left( \frac{1}{8}\left[({k_3}^2 - {k_4}^2)\left( 1 -
\cos(\sqrt{2}\omega t)\cosh(\sqrt{2}\omega t) \right) \right. \right. \nonumber \\
     &   & \left. \left. \qquad \qquad \qquad + \,2k_3 k_4
\sin(\sqrt{2}\omega t)\sinh(\sqrt{2}\omega t) \right] \right)^{1/3}\,, \label{eq26quater}
\end{eqnarray}
\begin{equation}
\varphi(t) = \log\left( \frac{B\left[ (k_3 - k_4)\sin(\sqrt{2}\omega t)\cosh(\sqrt{2}\omega t) + (k_3 +
k_4)\cos(\sqrt{2}\omega t)\sinh(\sqrt{2}\omega t) \right]}{A\left[ (k_3 + k_4)\sin(\sqrt{2}\omega
t)\cosh(\sqrt{2}\omega t) + (k_4 - k_3)\cos(\sqrt{2}\omega t)\sinh(\sqrt{2}\omega t) \right]}
\right)^{1/{\sigma}}\,. \label{eq26quinque}
\end{equation}
The scale factor in Eq. (\ref{eq26quater}) yields a sort of cyclic universe with spatially flat hyper-surfaces
(being $k = 0$). Since, in practice, it is nonvanishing only in an infinite number of limited intervals, this means
that the solution describing expansion should be considered only in one stage (in its first stage, for example).

On the other hand, the present value of the matter density parameter is given by
\begin{equation}
\Omega_{m0} \equiv 8\pi G D/({a_0}^3{H_0}^2) = 8\pi G D/{{\cal H}_0}^2 \sim 8\pi G D = 8\pi G \rho_{m0}\,,
\label{eq26six}
\end {equation}
being $a_0 = 1$ and $H_0\equiv {\cal H}_0 \sim 1$.

\subsection{The $\lambda > 0$ case}

Here, first of all, we have to distinguish between the two cases with $\epsilon = +1$ or $\epsilon = -1$, i.e.,
between Lagrangian functions in Eq. (\ref{eq16}) or Eq. (\ref{eq17}), respectively. Then, as we can see, our
considerations must also take the relative values of $AB$ and $\lambda$ into account. With respect to this, note
that the situation given by $A^2 = B^2 = 0$ describes a universe filled in with dust and free scalar field in
presence of a positive cosmological constant, since the potential $V(\varphi)$ reduces to $\lambda$ only. This
means we shall limit our considerations here to $AB\neq 0$.

\subsubsection{The $\epsilon = +1$ value}

First of all, let us note again that the case with $\epsilon = + 1$ and $\lambda = -2AB$ has been already
studied elsewhere \cite{de2} \cite{de2bis} \cite{de1}. Thus, we do not comment on it here. Now, on the other hand,
bearing in mind that $u^2 > v^2$ and $AB > 0$, the choice $\epsilon = +1$ always gives a positive potential for the scalar
field and leads to two different cases, since we can write (for $2AB \neq \lambda$)
\begin{equation}
\frac{{\sigma}^2}{2}(2AB + \lambda) \equiv {{\omega}_1}^2\,, \label{eq27}
\end{equation}
\begin{equation}
\frac{{\sigma}^2}{2}(2AB - \lambda) \equiv {\pm} {{\omega}_2}^2\,, \label{eq28}
\end{equation}
so that Eq. (\ref{eq16}) becomes
\begin{equation}
{\cal L}_{+} \equiv {\dot{u}}^2 - {\dot{v}}^2 + {{\omega}_1}^2u^2 {\pm} {{\omega}_2}^2v^2 + 2{\sigma}^2D\,.
\label{eq29}
\end{equation}
We thus deduce the equations
\begin{equation}
\ddot{u} = {{\omega}_1}^2u\,,\,\,\,\,\,\,\, \ddot{v} = {\pm} {{\omega}_2}^2v\,, \label{eq30}
\end{equation}
where the plus (minus) sign corresponds to the minus (plus) sign in the Lagrangian ${\cal L}_{+}$. In both cases
we find
\begin{equation}
u(t) = \alpha \exp{({\omega}_1t)} + \beta \exp{(- {\omega}_1t)}\,, \label{eq31}
\end{equation}
$\alpha$ and $\beta$ being integration constants. Furthermore, we find
\begin{equation}
v(t) = v_1 \sin{({\omega}_2t + v_2)} \label{eq32}
\end{equation}
when the plus sign ($2AB > \lambda$) is taken in the Lagrangian, or
\begin{equation}
v(t) = v_1 \exp{({\omega}_2t)} + v_2 \exp{(- {\omega}_2t)} \label{eq33}
\end{equation}
when the minus sign ($2AB < \lambda$) is chosen there. ($v_1$ and $v_2$ are integration constants.)

The integration constants are constrained by means of the equation $E_{{\cal L}_{+}} = 0$, which is
\begin{equation}
{\dot{u}}^2 - {\dot{v}}^2 - {{\omega}_1}^2u^2 {\pm} {{\omega}_2}^2v^2 - 2{\sigma}^2D = 0\,. \label{eq34}
\end{equation}
This gives a relationship that makes it possible to determine the more physically meaningful constant $D$ in terms of the
other constants, i.e.,
\begin{equation}
D = - \frac{\left({v_1}^2 {{\omega}_2}^2 + 4 \alpha \beta {{\omega}_1}^2\right)}{2{\sigma}^2} > 0 \label{eq35}
\end{equation}
when the minus sign is taken in Eq. (\ref{eq34}), and
\begin{equation}
D =  \frac{2(v_1 v_2 {{\omega}_2}^2 - \alpha \beta {{\omega}_1}^2)} {{\sigma}^2} > 0 \label{eq36}
\end{equation}
with the plus sign.
Focusing on the situation described by Eq. (\ref{eq32}), i.e. choosing $2AB >\lambda$, we see that the
inequality  (\ref{eq35}) implies that $- 4 \alpha \beta {{\omega}_1}^2
> {v_1}^2 {{\omega}_2}^2$, that is, $\alpha \beta < - {v_1}^2 (2AB - \lambda)/[4(2AB + \lambda)]$, giving
$\alpha \beta < 0$. So, when $\beta = - \alpha$, one finds $u(t) = 2 \alpha \sinh{({\omega}_1t)}$, and we may
write the volume $a^3(t)$ as the difference between the squares of a hyperbolic sine and a sine, where now $4
{\alpha}^2 {{\omega}_1}^2 > {v_1}^2 {{\omega}_2}^2$. On the other hand, choosing $v_2 = 0$ does not change the
sign of the value of $D$ and implies $a(0) = 0$, i.e., we can fix the origin of time as before, as well as in
Ref. \cite{rub2}. Of course, putting $v_1 = 0$ also leads to $a(0) = 0$, but then we always have $v(t) = 0$, so that $a^3(t) = 4{\alpha}^2 \sinh^2(\omega_1 t)$, being $u^2(0) = 0$. Anyway, let us stress that,
in the case under examination, it {\em must} be $a(0) = 0$ in order to avoid an initial negative value for the
scale factor.

At the same initial time $t=0$, the scalar field generally has the constant value
\begin{equation}
\varphi(0) = \frac{1}{\sigma} \log{\frac{B(\alpha + \beta + v_1 \sin v_2)}{A(\alpha + \beta - v_1 \sin v_2)}}\,,
\label{eq37}
\end{equation}
which gives rise to an undetermined form for $\beta = - \alpha$ and $v_2 = 0$, unlike what was found in
the case studied above, when $\lambda = 0$.

The cosmological solutions can be then written as
\begin{equation}
a(t) = \left[ \frac{4 {\alpha}^2 \sinh^2{({\omega}_1t)} - {v_1}^2 \sin^2{({\omega}_2t)}}{4} \right]^{1/3}\,,
\label{eq38}
\end{equation}
\begin{equation}
\varphi(t) = \frac{1}{\sigma} \log{\frac{B[2 \alpha \sinh{({\omega}_1t)}+ v_1 \sin{({\omega}_2t)}]}{A[2 \alpha
\sinh{({\omega}_1t)} - v_1 \sin{({\omega}_2t)}]}}\,, \label{eq39}
\end{equation}
so that the asymptotic behavior of the scale factor is an exponential one. As before, it thus expresses a
late--time {\em nonsoft} accelerated expansion of the universe, but a better agreement with observations demands
other considerations as to a more refined comparison with astrophysical data.

On the other hand, discussing the other allowed situation, with $2AB < \lambda$, the condition (\ref{eq36})
implies that $v_1v_2 {{\omega}_2}^2 > \alpha \beta {{\omega}_1}^2$, that is, $\alpha \beta < v_1v_2(\lambda -
2AB)/(2AB + \lambda)$, and now we can only say that $v_1v_2 < 0$ gives $\alpha \beta < 0$, while $v_1v_2 > 0$
leaves the possibility of having also $\alpha \beta > 0$. But let us note that $a(t) > 0$ involves $u - v > 0$
and, therefore, $u \neq 0$, which means that $\alpha \beta \neq 0$. If we then put $\beta = -\alpha$ (i.e.,
$\alpha \beta < 0$), we get $u(t) = 2 \alpha \sinh{{\omega}_1t}$, as before; this, in turn, requires that
${\alpha}^2 > - v_1v_2(\lambda - 2AB)/(2AB + \lambda)$, giving no restriction on the sign of $v_1v_2$, of
course.

Thus, we can also choose $v_2 = - v_1$ (implying ${\alpha}^2 > {v_1}^2(\lambda -2AB)/(2AB + \lambda)$) and get
$v(t) = 2 v_1 \sinh{({\omega}_2t)}$, so that we may write the volume $a^3(t)$ as the difference between the
squares of two hyperbolic sines. As before, this choice implies $a(0) = 0$ for every value of $\alpha$ and
$v_1$. Moreover, at $t = 0$,  the scalar field has in general the constant value
\begin{equation}
\varphi(0) = \frac{1}{\sigma} \log{\frac{B(\alpha + \beta + v_1 + v_2)} {A(\alpha + \beta - v_1 - v_2)}}\,,
\label{eq40}
\end{equation}
which is again undetermined when we choose $\beta = -\alpha$ and $v_2 = - v_1$.

Now, the cosmological solutions are written as
\begin{equation}
a(t) = \left[ {\alpha}^2 \sinh^2{({\omega}_1t)} - {v_1}^2 \sinh^2{({\omega}_2t)} \right]^{1/3}\,, \label{eq41}
\end{equation}
\begin{equation}
\varphi(t) = \frac{1}{\sigma} \log{\frac{B[\alpha \sinh{({\omega}_1t)} + v_1 \sinh{({\omega}_2t)}]}{A[\alpha
\sinh{({\omega}_1t)} - v_1 \sinh{({\omega}_2t)}]}}\,. \label{eq42}
\end{equation}
Of course, the expression of $a(t)$ is non negative only when the constants ${\alpha}$, ${\omega}_1$, $v_1$, and
${\omega}_2$ are such that the difference present in $a(t)$ is positive. In such a case, when $\omega_2 <
\omega_1$ we have an asymptotic exponential regime for the scale factor, as the one we already found before,
with the other choice for the signs of the constants. The case with $\omega_2 > \omega_1$, on the other hand,
has to be discarded since it never happens. It asymptotically leads to a non physical situation, with a negative
infinite value of the scale factor.

\subsubsection{The $\epsilon = -1$ value}

Let us consider, now, the Lagrangian in Eq. (\ref{eq17}), resulting from the potential $V_{-}$ expressed in Eq.
(\ref{eq15}). Since $V_{-}$ has to be positive, as we have chosen, it must be
\begin{equation}
4ABuv > - \lambda(u^2 - v^2)\,, \label{eq43}
\end{equation}
being $u^2 > v^2$, hence allowing $uv \gtrless 0$. (Note, in fact, that when $uv = 0$, the potential simply reduces
to $V_{-} = \lambda$, therefore becoming negative.)

Putting
\begin{equation}
{{\omega}_1}^2 \equiv {\sigma}^2 AB\,,\,\,\,\,\,\,\, {{\omega}_2}^2 \equiv \frac{1}{2}\lambda {\sigma}^2
\label{eq44}
\end{equation}
into Eq. (\ref{eq17}) gives
\begin{equation}
{\cal L}_{-} \equiv {\dot{u}}^2 - {\dot{v}}^2 + 2{{\omega}_1}^2uv + {{\omega}_2}^2(u^2 - v^2) + 2{\sigma}^2D\,,
\label{eq45}
\end{equation}
so that the resulting Euler--Lagrange equations are
\begin{equation}
\ddot{u} = {{\omega}_1}^2v + {{\omega}_2}^2u\,,\qquad \ddot{v} = - {{\omega}_1}^2u
+{{\omega}_2}^2v\,.\label{eq46}
\end{equation}

These equations constitute a linear system of homogeneous coupled ordinary differential equations of second
order, which can, more appropriately, be rewritten as:
\begin{equation}\label{sys1}
\left(
  \begin{array}{c}
   \ddot{u}(t) \\
    \ddot{v}(t) \\
  \end{array}
\right)=\left(
          \begin{array}{cc}
            a_{11} & a_{12} \\
            a_{21} & a_{22} \\
          \end{array}
        \right)\left(
                 \begin{array}{c}
                   u(t) \\
                   v(t) \\
                 \end{array}
               \right)\,,
\end{equation}
where $a_{11}=a_{22}=\omega_2^2$, and $a_{12}=\omega_1^2=-a_{21}$. The method of solving such a system reduces,
in fact, to diagonalizing the matrix
\begin{equation}
A=\left(
          \begin{array}{cc}
            a_{11} & a_{12} \\
            a_{21} & a_{22} \\
          \end{array}
        \right)\,.
\end{equation}
Actually, if we introduce the rotation matrix $\mathbb{R}$ (whose column are the eigenvectors
$\vec{\psi_1},\vec{\psi_2}$ of $A$), the transformation
\begin{eqnarray}\label{rotation}
&&\vec{y}= \mathbb{R} \vec{x}
\end{eqnarray}
(where we have posed $\vec{y}=\left(y_1(t),y_2(t)\right)$ and $\vec{x}=\left(u(t),v(t)\right)$) decouples our starting
system. One has
\begin{eqnarray}\label{rotation}
&&\ddot{\vec{ y}}= \mathbb{R} A \mathbb{R}^{-1} \vec{y}\,,
\end{eqnarray}
where $\mathbb{R} A \mathbb{R}^{-1}$ is the eigenvalues' diagonal matrix. Such eigenvalues are
\begin{eqnarray}
&&E_1=\frac{1}{2} \left(-\sqrt{(a_{11}-a_{22})^2 +4 a_{12} a_{21}}+a_{11}+a_{22}\right)\,,\\
&&E_2=\frac{1}{2} \left(\sqrt{(a_{11}-a_{22})^2 +4 a_{12} a_{21}}+a_{11}+a_{22}\right)\,,
\end{eqnarray}
and the eigenvectors associated with $E_{1}$ and $E_{2}$ are written as
\begin{eqnarray}
&&\vec{\psi_1}=-\frac{\left(\sqrt{(a_{11}-a_{22})^2+4 a_{12} a_{21}}-a_{11}+a_{22}\right)}{2 a_{21}}\hat{e_1}+ \hat{e_2}\,,\\
&&\vec{\psi_2}=\frac{\left(\sqrt{(a_{11}-a_{22})^2+4 a_{12} a_{21}}+a_{11}-a_{22}\right)}{2 a_{21}}\hat{e_1}+ \hat{e_2}\,.
\end{eqnarray}
In terms of the transformed vector $\vec{y}$, the system in Eq. (\ref{eq46}) reduces to a linear system of
homogeneous decoupled ordinary differential equations of second order, which can now be integrated exactly:
\begin{equation}\label{sys1}
\left(
  \begin{array}{c}
   \ddot{y}_1(t) \\
    \ddot{y}_2(t) \\
  \end{array}
\right)=\left(
          \begin{array}{cc}
            E_1 & 0\\
            0 & E_2\\
          \end{array}
        \right)\left(
                 \begin{array}{c}
                   y_1(t) \\
                   y_2(t) \\
                 \end{array}
               \right)\,.
\end{equation}

\begin{figure}
\centering{
        \includegraphics[width=8 cm, height=5 cm]{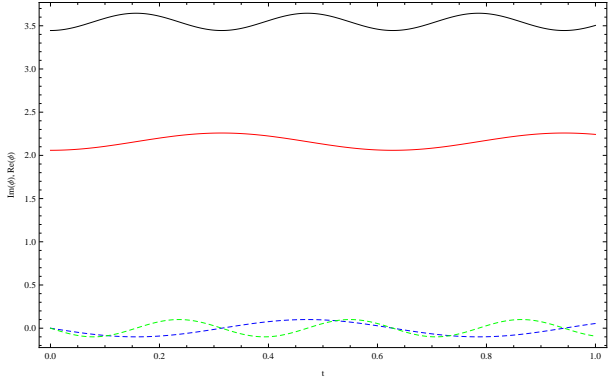}}
        \caption{Comparison between the time evolution of the real (solid lines) and the imaginary part (dashed lines)
        of the scalar field $\phi$, for different values of the integration constants. It turns out that the imaginary
        part of the scalar field slightly oscillates around zero, and the real part dominates. The solid and the dashed
        lines correspond to $c_2 = c_4=0 $\,, $c_3={\rm i} $\,, $c_1= 10 $\,\,, $\beta\equiv\arg \left(\omega _1^2+i
        \omega _2^2\right)=\frac{\pi}{3}$\,, $\omega_1= \frac{1}{\sqrt{2} n}$\,, and $\omega _2=
        \frac{\sqrt[4]{3}}{\sqrt{2} n}$. The solid red line and the dashed blue one correspond to $n=0.1$, while the
        others correspond to $n=0.05$.}
        \label{relimphi}
\end{figure}

The solution, in its general form, is
\begin{eqnarray}
&& y_1=c_1 e^{\sqrt{E_{1}} t}+c_2 e^{-\sqrt{E_{1}} t}\,,\\
&& y_2=c_3  e^{\sqrt{E_{2}} t}+c_4 e^{-\sqrt{E_{2}} t}\,.
\end{eqnarray}
The transformation $\mathbb{R}^{-1} \vec{y}$ provides the analytical expressions of $u(t)$ and $v(t)$:
\begin{eqnarray}
&& u(t)=-\frac{1}{2} i \left(c_1 e^{t \sqrt{\omega _1^2-i \omega _2^2}}+c_2 e^{-t \sqrt{\omega _1^2-i \omega
_2^2}}+ i e^{-t \sqrt{\omega _1^2+i \omega _2^2}}\left(c_4+c_3 e^{2 t \sqrt{\omega _1^2+i \omega
_2^2}}\right)\right)\,,
\label{u}\\
&& v(t)=\frac{1}{2} \left(i e^{-t \sqrt{\omega _1^2-i \omega _2^2}} \left(c_2+c_1 e^{2 t \sqrt{\omega _1^2-i
\omega _2^2}}\right)+c_3 e^{t \sqrt{\omega _1^2+i \omega_2^2}}+c_4 e^{-t \sqrt{\omega _1^2+i \omega
_2^2}}\right)\,,\label{v}
\end{eqnarray}
the $c_i$ being (where $i = 1....4$)  integration constants. It turns out that, in order to have a real-valued
function representing the scale factor $a(t)$, according to Eq. (\ref{eq12}), we have to set $c_2=c_4=0$,
$c_3=i$, so as to obtain the following form of the scale factor:
\begin{equation}\label{scalefac}
   a(t)= c_1 \frac{1}{4} e^{2 t \sqrt[4]{\omega _1^4+\omega _2^4} \cos \left(\frac{\beta }{2}\right)}\,.
\end{equation}
where we can set $c_1=\gamma\, a_0 $, $a_0$ being $ a(t_0)$ and $\gamma$ being a positive constant. With such a choice of
the integration constants we obtain a complex scalar field
\begin{equation}\label{eqphi}
   \phi(t)=\frac{i \sin \left(2 t \sqrt[4]{\omega _1^4+\omega _2^4} \sin \left(\frac{\beta }{2}\right)\right)}{c_1}-\frac{\cos \left(2 t \sqrt[4]{\omega _1^4+\omega _2^4}
   \sin \left(\frac{\beta }{2}\right)\right)}{c_1}-\frac{1}{2} \log \left(c_1^2\right)+\log \left(\omega _2^2\right)\,.
\end{equation}
Through an appropriate setting of the integration constants it is possible to see that the imaginary part of the
scalar field slightly oscillates around zero, and the real part dominates, as shown in Fig.\, (\ref{relimphi}).
In Eq. (\ref{scalefac}), on the other side, the scale factor exhibits an oscillation around an exponential
growth for large $t$, hence offering the possibility to describe the now observed stage of dark energy dominance.

\subsection{The $\lambda < 0$ case}

As above, we have again to discuss separately the two cases with $\epsilon = +1$ and $\epsilon = -1$, also
considering the relative value of $AB$ with respect to $\lambda$. First of all, let us note that, when $A^2 =
B^2 = 0$, we are once again in a universe with a free scalar field plus dust, but, now, in presence of a
negative cosmological constant. Thus, we shall once again limit our following discussion to the situation with
$AB\neq 0$, still describing the analysis as systematically as we can, in order to offer a useful tool in the whole
issue.

\subsubsection{The $\epsilon = +1$ value }

Starting from Eq. (\ref{eq16}), let us remember that we are examining the case with $AB > 0$, that is, when
${\sigma}^2(2AB + \lambda)/2 \equiv {\pm} {{\omega}_1}^2$ and ${\sigma}^2(2AB - \lambda)/2 \equiv
{{\omega}_2}^2$. This means that we have to consider two separate possible situations, due to the relative
values of $AB$ and $\lambda$.

The Lagrangian in Eq. (\ref{eq16}) is written as
\begin{equation}
{\cal L}_{+} \equiv {\dot{u}}^2 - {\dot{v}}^2 {\pm} {{\omega}_1}^2u^2 + {{\omega}_2}^2v^2 + 2{\sigma}^2D\,,
\label{min1}
\end{equation}
giving rise to the equations
\begin{equation}
\ddot{u} = {\pm} {{\omega}_1}^2 u\,, \qquad \ddot{v} = - {{\omega}_2}^2 v\,, \label{min2}
\end{equation}
where the plus (minus) sign corresponds to the minus (plus) in Eq. (\ref{min1}). The solution for $u(t)$ is
\begin{equation}
u(t) = u_1 \sin ({\omega}_1 t + u_2) \label{min3}
\end{equation}
for the minus sign in Eq. (\ref{min1}) (i.e., with $2AB < \lambda$), and
\begin{equation}
u(t) = u_1 \exp ({\omega}_1 t) + u_2 \exp (- {\omega}_1 t) \label{min4}
\end{equation}
for the plus sign (i.e., with $2AB > \lambda$). ($u_1$ and $u_2$ are integration constants.) The solution $v(t)$
is in both cases
\begin{equation}
v(t) = \alpha \sin ({\omega}_2 t + \beta)\,, \label{min5}
\end{equation}
$\alpha$ and $\beta$ being integration constants.

The constraint equation $E_{{\cal L}_{+}} = 0$ now becomes
\begin{equation}
{\dot{u}}^2 - {\dot{v}}^2 \pm {{\omega}_1}^2u^2 - {{\omega}_2}^2v^2 - 2{\sigma}^2D = 0\,, \label{min6}
\end{equation}
leading to
\begin{equation}
D = - \frac{\left({\alpha}^2 {{\omega}_2}^2 + 4u_1 u_2 {{\omega}_1}^2\right)}{2{\sigma}^2} > 0 \label{min7}
\end{equation}
for the plus sign in ${\cal L}_{+}$ (corresponding to the minus one in Eq. (\ref{min6})), and to
\begin{equation}
D = - \frac{\left({u_1}^2 {{\omega}_1}^2 - {\alpha}^2 {{\omega}_2}^2\right)}{2{\sigma}^2} > 0 \label{min8}
\end{equation}
for the alternative case with the minus sign in ${\cal L}_{+}$ (and the plus sign in Eq. (\ref{min6})).

From (\ref{min7}), we must have $u_1 u_2 < - {\alpha}^2 (2AB - \lambda)/[4(2AB + \lambda)] <
0$. Now, choosing $u_2 = - u_1$, one finds $u(t) = 2u_1 \sinh ({\omega}_1 t)$, so that the volume $a^3 (t)$ is
written as the difference $(4{u_1}^2 \sinh^2 ({\omega}_1 t) - {\alpha}^2 \sin^2 ({\omega}_2 t + \beta))/4$, with
${u_1}^2 > {\alpha}^2 (2AB - \lambda)/[4(2AB + \lambda)]$. If we then set $\beta = 0$, this is consistent with
the physically required positive sign of $D$ and, at the same time, annihilates $a(0)$. Note, on the other side,
that the value $\alpha = 0$ gives $v(t) = 0$ at any $t$, and $a^3 (t) = {u_1}^2 \sinh^2 ({\omega}_1 t)$, so that
this choice still yields $a(0) = 0$. If we do want to have both $a(0) = 0$ and assume $\alpha = 0$ (leading to a
finite constant $\varphi(0)$as well as -- as we already said--, as an initially vanishing scale factor), we instead (and
unfortunately) find a constant scalar field at any time. Thus, even if it is true that, when both $u_2 = - u_1$
and $\beta = 0$, the argument of the logarithm in the $\varphi$--field becomes an undetermined form, the condition
$a(0) = 0$ does require $u_2 = - u_1$ and $\beta = 0$.

With such choices the cosmological scale factor and the scalar field are
\begin{equation}
a(t) = \left( \frac{4{u_1}^2 \sinh^2 ({\omega}_1 t) - {\alpha}^2 \sin^2 ({\omega}_2 t)}{4} \right)^{1/3}\,,
\label{min9}
\end{equation}
\begin{equation}
\varphi(t) = \frac{1}{\sigma} \log \frac{B(2u_1 \sinh (\omega_1 t) + \alpha \sin (\omega_2 t))}{A(2u_1 \sinh
(\omega_1 t) - \alpha \sin (\omega_2 t))}\,. \label{min10}
\end{equation}
But, even if this solution would be asymptotically good for describing an inflationary stage at present in the
universe, we must rule it out because of the above considerations..

Turning, thus, to the situation leading to Eq. (\ref{min8}), we see that it has to be ${u_1}^2 < {\alpha}^2
({\omega_2}^2/{\omega_1}^2)$ in order to get $D > 0$. This does not imply special restrictions on the constants
involved, and, without making any particular choice for them, we can now write the cosmological solutions as
\begin{equation}
a(t) = \left( \frac{{u_1}^2 \sin^2 ({\omega}_1 t + u_2) - {\alpha}^2 \sin^2 ({\omega}_2 t + \beta)}{4}
\right)^{1/3}\,, \label{min11}
\end{equation}
\begin{equation}
\varphi(t) = \frac{1}{\sigma} \log \frac{B(u_1 \sin (\omega_1 t + u_2) + \alpha \sin (\omega_2 t +
\beta))}{A(u_1 \sin (\omega_1 t + u_2) - \alpha \sin (\omega_2 t + \beta))}\,. \label{min12}
\end{equation}

As to the time behavior of the scale factor, we here limit ourselves to note that it has to be
\begin{equation}
a(0) = \frac{{u_1}^2 \sin^2 (u_2) - {\alpha}^2 \sin^2 (\beta)}{2^{2/3}} = 0\,,
\end{equation}
which soon involves a constraint on the integration constants, that is, ${u_1}^2 \sin^2 (u_2) = {\alpha}^2
\sin^2 (\beta)$. Putting $u_2 = \beta = 0$ in Eq. (\ref{min11}), we soon get $a(0) = 0$. On the other hand, a
choice like $u_2 = \alpha = 0$ also gives the same result. Furthermore, these two possibilities both give a
finite constant initial value for $\varphi(0)$. Afterwards, the scale factor evolves oscillating with time,
describing alternate phases of closed and open universes. Whether this can be the case for the universe we live
in, however, remains an open problem that needs further investigation.

\subsubsection{The $\epsilon = -1$ value }

This situation has to be considered, now, starting from the Lagrangian in Eq. (\ref{eq17}). Putting ${\omega}^2
\equiv {\sigma}^2 AB$ then yields
\begin{equation}
{\cal L}_{-} = {\dot{u}}^2 - {\dot{v}}^2 + 2 {\omega}^2 u v + \frac{1}{2} \lambda {\sigma}^2 (u^2 - v^2) + 2
{\sigma}^2 D\,, \label{min13}
\end{equation}
from which we deduce the equations
\begin{equation}
\ddot{u} = \frac{1}{2} \lambda {\sigma}^2 u + {\omega}^2 v\,, \qquad \ddot{v} = \frac{1}{2} \lambda {\sigma}^2 v
- {\omega}^2 u\,. \label{min14}
\end{equation}

It turns out that this system is of the same kind as the one in Eqs. (\ref{eq46}), with
$a_{11}=a_{22}=\frac{1}{2} \lambda {\sigma}^2$\,, and $a_{12}=-a_{21}= {\omega}^2$, so that the solutions can be
represented as in Eq. (\ref{scalefac}) and Eq. (\ref{eqphi}).

\section{Connection between quintessence and inflation for exponential potential models}

During inflation, the scalar field $\varphi$ is supposed to be highly excited and slowly evolving (``rolling
down'') to the minimum of the potential. In this phase the potential energy of the inflaton field dominates the
energy density of the universe. After the period of inflation $\varphi$ rapidly oscillates around the minimum of
the potential and, by virtue of coupling with other scalar and spinor matter fields, massive and massless particles
are created and the universe reheats, starting the standard post-Big-Bang evolution. Simple models of inflation
use a scalar field with a potential of the form $V(\varphi)={\frac{1}{2}}m^{2}\varphi^{2}$,
where $m$ is the mass of the inflaton field. In the standard scenario the energy of the inflaton field is then
transformed into mass--energy of created particles. Another possibility is offered by a quartic potential
\begin{equation}\label{potential1}
W(\varphi)=\alpha\left(\varphi^2 -\delta^2\right)^2,
\end{equation}
where $\alpha$ and $\delta$ are parameters.

Many efforts have been made for constructing unified frameworks for inflation and quintessence which employ a
unique scalar field to drive both stages (see, for instance, Refs. \cite{peebles} \cite{neupane} \cite{ bastero-gil}
\cite{dimopoulos}). Actually, in such scenarios the scalar field responsible of late time acceleration is
nothing else but the remnant of the one which caused inflation at early time. Indeed, a successful model  of
quintessential inflation is subject to the constraints of both inflation and quintessence simultaneously. For
example, the minimum of the potential must not have been yet reached by the scalar field (generally, this
requirement is satisfied by assuming the presence of a quintessential tail, that is assuming potentials with the
minimum displaced at infinity). Moreover, it is needed that $\varphi$ should not decay completely into a thermal
bath of particles in order to survive until today, just to drive the late phase of accelerating expansion. As a
consequence of this, the universe undergoes a period of \textit{kination expansion}, when its energy density is dominated
by the kinetic energy of $\varphi$. In this context, the standard reheating mechanism usually assumed to
generate the primordial plasma does not work; however, the mechanism of gravitational particle production can
still reheat the universe in the framework of quintessential inflation. Actually, even if the amount of
radiation produced by such a mechanism is largely sub--dominant when compared with the energy field contribution
during kination, the energy density of kination is redshifted by the cosmological expansion much faster than the
radiation density and will start to dominate at some temperature.

Generally, the constraints and requirements which are to be satisfied  by quintessential inflation are satisfied
by using a multi--branch scalar potential, where the change of the potential, when the field moves from the
inflationary to the quintessential frame of its evolution, is fixed by hand \cite{peebles} or is the outcome of
a phase transition arranged by the interaction with other scalar fields (see, for instance, Ref. \cite{K-R}).
Recently, the possibility has been investigated to connect the inflationary and quintessential expansion of
the universe within the theoretical framework of particle production, usually developed in the very early
universe. (Even if the quantum aspect of the creation mechanism is not yet very well known, some classical
aspects due to kinetic collision in the hot dense regions of the early universe have been discussed in the
literature.) This suggests that the same mechanism may occur in the late universe, also leading to late time
cosmic acceleration (see, for instance, Refs. \cite{lima} \cite{sanyal}). In other approaches, a large variety
of quintessential inflationary potentials are derived from theories of non--minimally coupled gravity (see, for
instance, Ref. \cite{kaganovich}). In our model the quintessential inflation is formulated in terms of a
multi--branch scalar field, driving both the inflationary and the quintessential phases of the evolution of the
universe. The quintessential tail is realized through an exponential potential (hence choosing the latter as our
working potential for quintessence); on the contrary, to describe the inflationary plateau we do not fix any
inflationary potential, but we propose a parametrization of the inflationary scalar field equation of state, and
implement the transition from an inflationary stage to a kination evolution, which is characterized by the value
$w_\varphi=1$ of the equation of state, and corresponds just to the asymptotic in the past value for the
equation of state of the exponential potential scalar field.

\subsection{The fiducial cosmological model}

In order to illustrate our paradigm, we use, as fiducial cosmological model, the one considered in Refs.
\cite{rub3} \cite{dem}. This model is based on the simplest form of exponential potential of the quintessence
field
\begin{equation}\label{potexp}
V(\varphi)= V_{0}e^{-\sqrt{3\over 2}\varphi}\,,
\end{equation}
and the assumption that the universe is spatially flat and filled in with dark matter and scalar field. We
postpone to a forthcoming paper the detailed investigation of the impact of a kination--dominated phase generated
by our class of potentials. The equations that determine the dynamics of this model are the Friedmann equation
\begin{equation}\label{friedmann}
 3H^{2}=\varrho_{m}+\varrho_{\varphi}
\end{equation}
(where we use units in which $8\pi G=c=1$, $H={{\dot a}\over a}$ is the Hubble constant, $\varrho_{m}\sim
a^{-3}$ is the density of matter, and $\varrho_{\varphi}={1\over 2}{\dot \varphi}^{2}+V(\varphi)$ is the energy
density of the scalar field), the Raychaudhury equation
\begin{equation}\label{raych}
2{\dot H} + 3H^{2}= -({1\over 2}{\dot \varphi}^{2}-V(\varphi))\,,
\end{equation}
and the generalized Klein--Gordon equation describing the evolution of the scalar field
\begin{equation}\label{KG}
{\ddot \varphi}+3H{\dot \varphi}+{{dV}\over {d\varphi}}=0\,.
\end{equation}

\begin{figure}
\centering{
        \includegraphics[width=8 cm, height=6 cm]{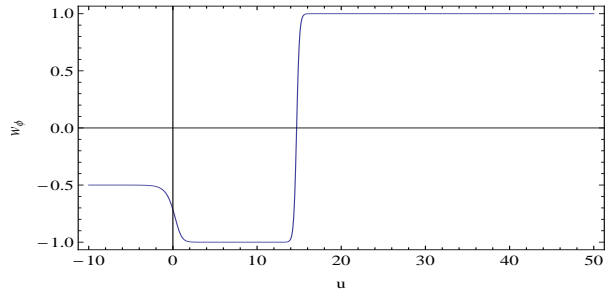}}
        \caption{The dark energy equation of state $w_{\varphi}$ parameter as a function of $u$.}
        \label{wm}
\end{figure}

\begin{figure}
\centering{
        \includegraphics[width=8 cm, height=5 cm]{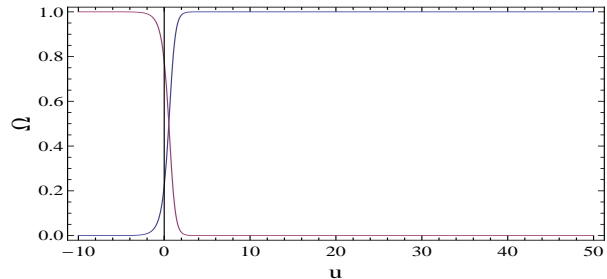}}
        \caption{Omega parameters as functions of $u$, where $\Omega_{\varphi}$ is marked
        in green and $\Omega_{m}$ in blue.}
        \label{omegam}
\end{figure}

These equations get a simple form when instead of $t$ as an independent variable one uses $a(t)$, the scale
factor. Introducing a new independent variable by $u=\log(1+z)=-\log({a(t)\over a_{0}})$, where $a_{0}$ is the
present value of the scale factor and $z$ is the redshift, and rescaling other variables as ${\bar
\varphi}={1\over \sqrt{3}}\varphi$, ${\bar \varrho_{i}}= {\varrho_{i}\over {3H_{0}^{2}}}$ (where $i = m,
\varphi$), ${\bar V_{0}}={V_{0}\over {3H_{0}^{2}}}$ and ${\bar H}={H\over H_{0}}$, where $V_0$ is a parameter
and $H_{0}$ the present value of the Hubble constant, we thus obtain the following set of equations that contain
only dimensionless variables
\begin{equation}
{\bar H}^{2}= {{{\bar \varrho_{m}}+{\bar V}}\over {1- {1\over 2}{{\bar\varphi}}'^{2}}}\,,
\end{equation}
\begin{equation}
{\bar H}^{2}{\bar \varphi}'' - 3({1\over 2}{\bar \varrho_{m}}+{\bar V}){\bar \varphi}'+{{d{\bar V}}\over {d{\bar
\varphi}}}=0\,,
\end{equation}
with prime denoting derivative with respect to $u$. Such equations can be solved analytically but we here limit
ourselves only to present the plots of some important quantities of the model, such as $w_{\varphi}(u)$,
$\Omega_{m}(u)$ and $\Omega_{\varphi}(u)$. (See Fig. \ref{wm} and Fig. \ref{omegam} for their behaviors.)

This model of the universe is described by an exact solution of the dynamical equations. The arbitrary
parameters that appear in the solution are determined by specifying the initial conditions; we in fact require
that at the present time, e.g. at $u=0$, one should have $\Omega_{m}(u=0)=0.3$ and $\Omega_{\varphi}(u=0)=0.7$. The
variable $u$ is such that it decreases as time increases, and, at early times, the scalar field is almost
constant and only recently it starts to increase.  On the other hand, the potential of the scalar field at early
times is constant and only recently is rapidly decreasing, while in the future it assumes a constant value
again. The dark energy equation of state $w_{\varphi}$ parameter in the far past is equal to $ -1$, so that at
the early stage of the evolution of the universe dark energy behaves as a cosmological constant, but (as is
shown in Fig. \ref{omegam}, through the behaviors of the $\Omega$ parameters) it only recently has started to
dominate the expansion rate of the universe.

The quintessential exponential potential admits, as said, exact solutions of the Einstein field equations.
Actually, from them we find \cite{rub2}
\begin{equation}\label{varphi}
\varphi=-\sqrt{\frac{2}{3}}\log\left(  \frac{2}{1+t^{2}}\right)\,,
\end{equation}
which was obtained using as suitable unit of time the age of the universe, i.e. $t_{0}=1$. Thus, at the present
time $\varphi_{0}=0$. At the time of reheating $t$ is virtually zero, so that we may set $\varphi_{in}
\equiv -\sqrt{\frac{2}{3}\log \left(  2\right)}$. We now shift $\varphi$ to $\ x \equiv \varphi-\varphi_{in}$,
so that the potential may be written as
\begin{equation}\label{pot}
V=4\exp\left(  -\sqrt{\frac{3}{2}}x\right)\,,
\end{equation}
where the prefactor $4$ is due to the particular choice of units (and, of course, is also the initial value for
$V$) and represents the value of the effective cosmological constant at that time, so that it has the dimension
of a $[{\rm length}]^{-1}$. In our units, the unit length is of the order of Hubble length, just as the current
estimate for $\Lambda$. An effective cosmological constant near to $4$ is thus slightly greater than this and
slowly evolves towards a smaller value of $2$ nowadays.

Let us now consider more realistically the inclusion of radiation, too, into such a model. In this case the
dynamical equations as far as we know do not have analytical solutions, and therefore we will rely on numerical
computations. Following the procedure used above, we use the variable $u$ instead of time $t$ together with the
rescaled variables, so that the equations contain only dimensionless variables and can now be written in the
form
\begin{equation}
{\bar H}^{2}= {{{\bar \varrho_{m}}+{\bar \varrho_{r}}+{\bar V}}\over {1- {1\over 2}{{\bar \varphi}}'^{2}}}\,,
\end{equation}
\begin{equation}
{\bar H}^{2}{\bar \varphi}{''} - ({\bar \varrho_{r}}+{3\over 2}{\bar \varrho_{m}}+3{\bar V}){\bar
\varphi}'+{{d{\bar V}}\over {d{\bar \varphi}}}=0\,,
\end{equation}
where ${\bar\varrho_{r}}\sim a^{-4}$ is the rescaled energy density of radiation.

\begin{figure}
\centering{
        \includegraphics[width=8 cm, height=5 cm]{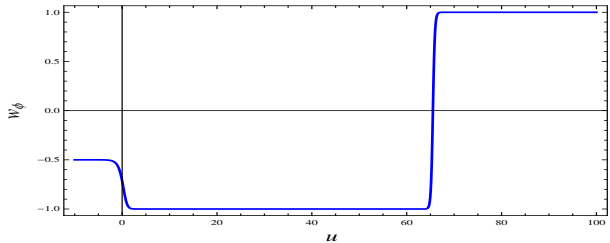}}
        \caption{The dark energy equation of state $w_{\varphi}$ parameter as a function of $u$ with radiation
        included.}
        \label{wr}
\end{figure}

\begin{figure}
\centering{
        \includegraphics[width=8 cm, height=5 cm]{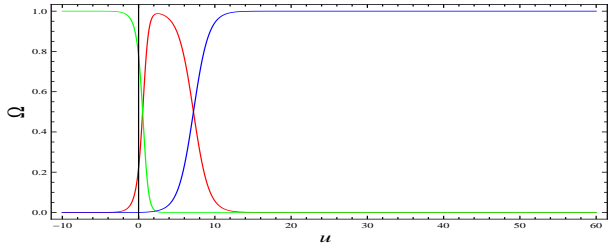}}
        \caption{Omega parameters as a function of $u$ in the universe filled in with matter, radiation and scalar
        field. $\Omega_{\varphi}$ is marked in green, $\Omega_{r}$ in red and $\Omega_{m}$ in blue.}
        \label{omegar}
\end{figure}

We can numerically solve this system of coupled equations, specifying the initial condition at $u=60$, for
example, and assuming that $\varphi(30)$, ${\varphi'(30)}$, and $H(30)$ have the same values as in the case
without radiation, while $\Omega_{r}(60)$ is just the rescaled present value of $\Omega_{r}$. Some results of
numerical integration are shown in Figs. \ref{wr} and \ref{omegar}. It can be seen that the presence of
radiation is slightly changing the behavior of the scalar field, its potential, the Hubble constant, and the $w$
parameter of the dark energy equation of state. As expected, now, only the evolution of the $\Omega$ parameters
is different. At the initial time, again when $u=30$, radiation dominates the expansion rate of the universe,
with dark energy and matter being subdominant, at a redshift $z$ of about $5000$; the energy densities of matter
and radiation become equal and, for a relatively short period, the universe becomes matter dominated until, at a
redshift of about $1$, dark energy starts dominating the expansion rate of the universe. With these results we
then confirm the independent investigations of effects of radiation on the evolution of the quintessence field
by Urbano and Rosenfeld (\cite{franca}). From our results it indeed follows that, during the epoch of
nucleosynthesis ($z\sim 10^{9}$),  the energy density of the scalar field is much smaller than the energy density
of radiation. In particular, during such an epoch the kinetic term in the scalar field energy density vanishes,
and the potential term is constant, so that the dark--energy term acts as an effective cosmological constant
$\Lambda$ and does not affect the process of primordial nucleosynthesis.

\subsection{Considerations about the inflation--\textit{kination} transition}

On the other side, as remarked above, an indispensable ingredient of quintessential inflationary scenarios is
the existence of an early kination--dominated (KD) era, where the universe is dominated by the kinetic energy of
the quintessence field. During this era, the expansion rate of the universe is larger compared to its value
during the usual radiation--domination (RD) epoch. A generic reasonable inflation--kination period can usually
be described by two main parametrizations of the transition, without specifying the details of the modelization.
One is a polynomial parametrization for $a^2(\eta)$ ($\eta$ being the conformal time) \cite{ford}, and the other
is a parametrization of the equation of state $w$, smoothly connecting the inflation ($w=-1$) and kination
($w=1$) regimes. (See, for instance, Ref. \cite{chun}, where a hyperbolic tangent parametrization is used to
study gravitational reheating in quintessential inflation). Moreover, the impact of a kination--dominated phase
generated by a quintessential exponential model or by quintessential power law, or also by a running kinetic
inflation model, has been already investigated in literature (see for instance \cite{gomez},\cite{chun},
\cite{nakayama}). Here our approach is quite different: actually, we are not fixing a single--branch scalar
field potential which drives both the early--time inflationary and the late--time quintessence evolution of the
universe; we instead want to show that a quintessential exponential potential tail can be naturally connected to
the inflation through a kination dominated era, since the  \textit{asymptotic} value (in the past) of the
equation of state, $w_\varphi$, is just $w_\varphi=1$, which characterizes the kination phase. We illustrate
such a mechanism by using a new parametrization of the scalar field equation of state, which could also be used to
study statistical properties of massive non relativistic bosons arising at the first stage of reheating as a
result of a quantum decay of a classical quantum field in inflationary cosmological model without slow rolling
\cite{deritis95}.

The FRW evolution of the universe in the inflationary epoch is described by the equations
\begin{eqnarray}
&&3H^2 = \rho_\varphi \,,\label{eq1infla}\\
&&\frac{\ddot{a}}{a} = -\frac{\rho_\varphi+3 P_\varphi}{2}\,,\label{eq2infla}
\end{eqnarray}
where we are using dimensionless units and $a_0= \sqrt{\frac{3}{8\pi}}l_{Pl}\frac{E_{Pl}}{E_{GUT}}^2\simeq
10^{-25}\, cm$. The quantities $\rho_\phi$ and $P_\phi$ are related by the equation of state $P_\varphi=w_\varphi
\rho_\varphi=\left(\gamma_\varphi-1\right)\rho_\varphi$, and their evolution is described by the continuity
equation
\begin{equation}\label{eq3infla}
\dot{\rho_\varphi}+3 H (\rho_\varphi+P_\varphi)=0\,.
\end{equation}
From Eqs. (\ref{eq1infla}), (\ref{eq2infla}), and (\ref{eq3infla}) it turns out that
\begin{equation}
\gamma_\varphi=-\frac{2}{3}\frac{\dot{H}}{H^2}\,.\label{eq4infla}
\end{equation}
It is straightforward to see that for $\gamma_\varphi=0$ we have an exponential behavior for $a(t)$; if
$\gamma_\varphi=\gamma_0$ is constant, there is a power-law expansion, and we have inflation iff $\gamma_0<
\frac{2}{3}$,  the more de Sitter--like, the closer $\gamma_0$ to zero.

All this suggests parametrizing the dynamics of the universe during such a phase in terms of
$\gamma_\varphi(t)$. Indeed, we have that
\begin{eqnarray}
H(t) &=& \frac{2}{3}\left(\int \gamma_\varphi dt\right)^{-1}\label{gamma1}\,,\\
\rho_\varphi(t) &=& \frac{4}{9}\left(\int \gamma_\varphi dt\right)^{-2}\label{gamma2}\,, \\
P_\varphi &=& \frac{4}{9}\left(\gamma_\varphi-1\right)\left(\int \gamma_\varphi dt\right)^{-2}\label{gamma3}\,, \\
V(\varphi(t)) &=& H^2(t) \left(1-\frac{1}{2}\gamma_\varphi\right)\label{gamma4}\,,\\
\varphi(t) &=& \int \sqrt{-\frac{2}{3}\dot{H}}dt\label{gamma5}\,.
\end{eqnarray}
The above Eqs. (\ref{gamma1}) and (\ref{gamma5}) are parametric equations for the potential $V$, and can allow
to reconstruct the scalar field potential. Thus, let us consider a scalar field which at the beginning describes
an inflationary stage, and undergoes a phase transition into a kination phase. This can be achieved by postulating
a special form for $\gamma_\varphi$ (and, therefore, $w_\varphi$), and  reconstructing the potential.

\begin{figure}
\centering{
        \includegraphics[width=8cm, height=5 cm]{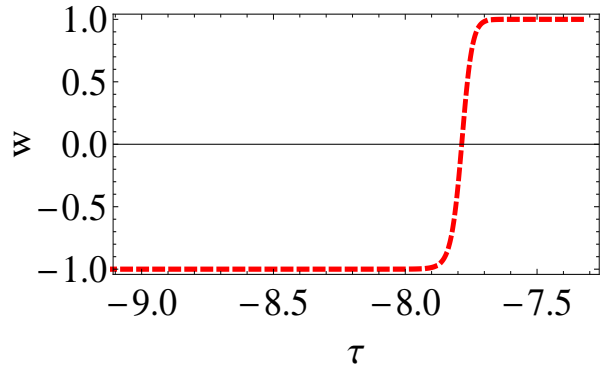}}
        \caption{Equation of state $w_\varphi$ as a function of time, when the universe is dominated by  kination
        at the end of inflation. The parametrization of Eq. (\ref{gammaform}) is used, with $\alpha=2$, $\beta=1.32$,
        $t_f=16.9$.}
        \label{wquintinfla}
\end{figure}

\begin{figure}
\centering{
        \includegraphics[width=8cm, height=5 cm]{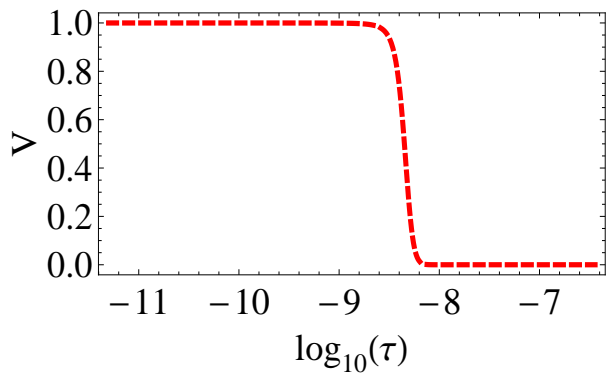}}
        \caption{Behavior of the scalar field potential as a function of time, when the universe is dominated by
        kination at the end of inflation. The parametric equations for the potential $V$ in  Eq. (\ref{gamma4}) is
        used, with $\gamma_\varphi$ given in Eq. (\ref{gammaform}) and $\alpha=2$, $\beta=1.32$, $t_f=16.9$.}
        \label{Vquintinfla}
\end{figure}

\begin{figure}
\centering{
        \includegraphics[width=8cm, height=5 cm]{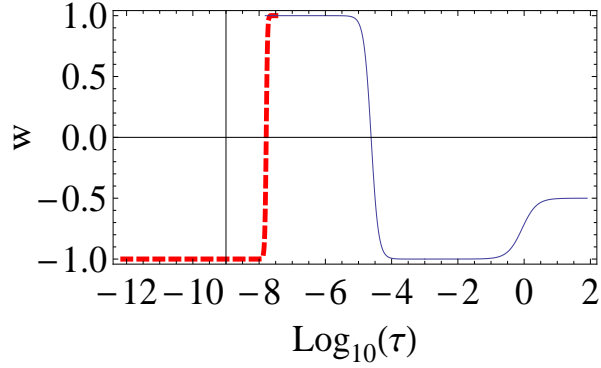}}
        \caption{Evolution of the equation of state from the inflationary to the quintessential stage. Dashed in red
        the transition  is plotted from the inflation into the kination when  $w_\varphi=1$; we se that such an
        \textit{asymptotic in the future} value for the equation of state corresponds to the \textit{asymptotic in the
        past} value for the quintessential exponential potential (solid line).}
        \label{wquintinfla2}
\end{figure}

\begin{figure}
\centering{
        \includegraphics[width=8cm, height=5 cm]{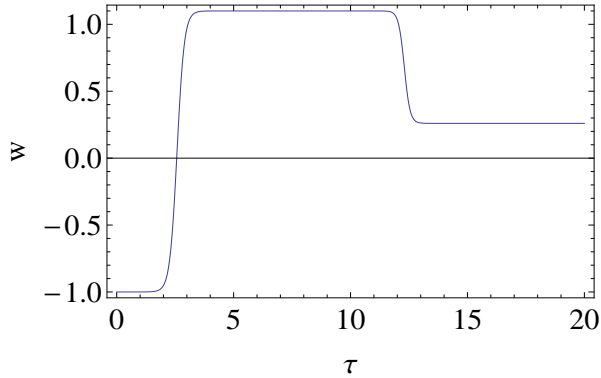}}
        \caption{ We show the double transition from vacuum into kination and into radiation obtained by using the
        parametrization of Eq. (\ref{newpar}), with  $A=2.1$\,,$B=0.84$\,\,\,$\beta=8.14$\,,$\alpha =-6.43$\,,
        $t_f=12.3$\,, $t_0=2.58$.}
        \label{wquintinflaradiation}
\end{figure}

\begin{figure}
\centering{
        \includegraphics[width=8cm, height=5 cm]{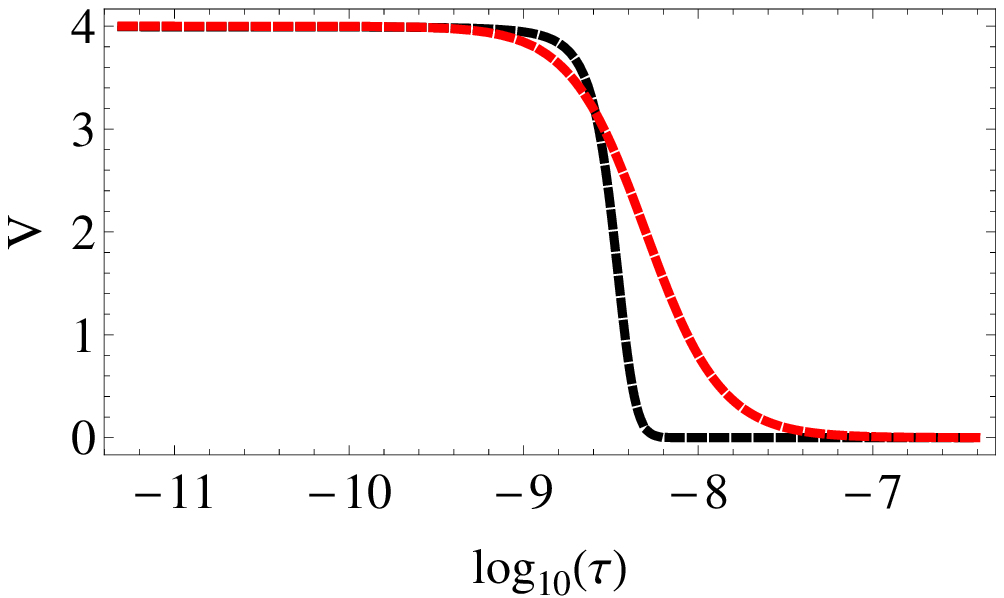}}
        \caption{  Behaviour of the scalar field potential during the two stages: it turns out that the quintessential
        exponential potential (red dashed line) asymptotically behaves like the inflationary one (black dashed line).}
        \label{vcompare}
\end{figure}

Our choice for $\gamma_\varphi$ is
\begin{equation}\label{gammaform}
\gamma_\varphi = \alpha \frac{1}{1+\frac{1}{2}\exp(-\beta (t-t_f))}\,,
\end{equation}
where $\beta$ and $t_f$ are parameters giving, respectively, the rate of the phase transition and the time at
which inflation ends. On the other hand, the parameter $\alpha$ is related to the
\textit{asymptotic} value of $w_\varphi$, which in our model is $w_\varphi=1$. (For different values of
$\alpha$, it is possible to obtain a transition from vacuum into dust or radiation. Moreover, by using a slightly
different parametrization of the equation of state, as
\begin{equation}
\gamma(\varphi)=\frac{A}{e^{\alpha  (t-t_0)}+1}-\frac{B}{e^{\beta  (-(t-t_f))}+1}-1\,, \label{newpar}
\end{equation})
it is also possible to obtain a double transition from vacuum into kination and into radiation, as illustrated
in Fig. (\ref{wquintinflaradiation}). Such a form, even if arbitrarily assigned, can easily implement the
transition from an inflationary stage to a kination one, as we see in Figs. \ref{wquintinfla} and
\ref{Vquintinfla}, where we plot the behavior of the equation of state and the reconstructed scalar field
potential, respectively. We have to note that the potential changes rather roughly from its value during
inflation to its final values, and that the \textit{asymptotic in the future} value for the equation of state
$w_\varphi=1$ corresponds to the \textit{asymptotic in the past} value for the quintessential exponential
potential. Thus, these behaviors can be naturally connected, as shown in Fig. \ref{wquintinfla2}, where the
time--scale has been arbitrarily taken in order to show all the evolution from inflation into kination, toward
the late--time exponential quintessential stage, and the values of the parameters $\alpha$, $\beta$ and $t_f$
are fixed as just optimizing the link between the early and late time evolutions. Finally, in Fig. \ref{vcompare}
we compare the behavior of the scalar field potential (with radiation included) during the two stages: it turns
out that the quintessential exponential potential asymptotically behaves like the inflationary one.


\section{Discussion and conclusions}
After the historical role of scalar fields in inflationary cosmology, the quite recent reconsideration of their
crucial importance relies upon the fact that they can improve a dynamical mechanism for giving rise today to a
repulsive component in cosmic energy, the dark energy. The interest for a scalar field in the universe,
otherwise usually thought of as a matter--dominated one, is in fact tied to the very recent cosmological history
revealed by astrophysical observations.

In this paper, our attention has been generally focused on the role the exponential--like potential has still to
play in nowadays cosmology. At first, we have discussed some models generalizing the simple exponential
form of the potential, in order to derive general exact solutions. The technique used has been very simple, being it based on a given {\em ad hoc} change of variables. Such a procedure is generally possible either \emph{by chance} or because there exists a sort of a method to deduce that useful transformation. In our work we have adopted the second procedure, i.e. the
Noether Symmetry Approach to cosmology, borrowing from it here not only the suggestion on the kind of useful
transformations to apply to the variables $a$ and $\varphi$ involved, but also on the \emph{natural} kind of
potential $V(\varphi)$ to study. On the other hand, the results we have found in this first part of the paper
are many and not always easily discernable. Basically, they have been mathematically derived but not yet
appropriately discussed on physical ground. This deserves, of course, further investigations in a forthcoming
paper, but we can however try here to sum up what has come out as more interesting for nowadays cosmology.

When $\lambda = 0$ in the potential, first of all, the only case we have discussed here is for $A^2 \neq 0\,,
B^2 \neq 0\,,\epsilon = - 1$. The universe then displays infinitely many periods of closed evolution, and one should
choose which branch to investigate, anyway without succeeding in describing the currently observed behaviour. The
situation with $\lambda > 0$, on the other hand, depends on the sign of $\epsilon$, even if it presents
accelerated asymptotic evolutions both for $\epsilon = + 1$ and $\epsilon = - 1$. In the latter case, such an
acceleration is however obtained together with an oscillating behavior. This is also true for $\lambda < 0$ when
$\epsilon = - 1$, while for $\epsilon = + 1$ we again find a closed universe.

It remains to be seen that the only physically acceptable situation (and solution) is given for $\lambda > 0$
and $\epsilon = + 1$. In a forthcoming paper we are going into details of the cosmological evolution, with regard to each specific solution discussed above. In particular, we are going to perform the
necessary confrontation of this theoretical output with the observational data sets.

The second part of the paper is still connected with the exponential potential. But, now, we instead
focus on considering and illustrating a possible quintessential inflationary scenario, formulated in terms of a
multi--branch scalar field, driving both the inflationary and the quintessential phases of the evolution of the
universe. The quintessential tail is realized through an exponential potential (choosing, thus, the latter as our
working potential for quintessence); on the contrary, for describing the inflationary plateau we do not
fix any inflationary potential, but we propose a parametrization of the inflationary scalar field equation of
state, and implement the transition from an inflationary stage to a kination evolution, which is characterized
by the value $w_\varphi=1$ of the equation of state, which is needed in any reasonable quintessential
inflationary model and corresponds just to the asymptotic in the past value for the equation of state of the
exponential potential scalar field. It turns out that the reconstructed potential changes rather roughly from
its value during inflation to its final values, and the \textit{asymptotic in the future} value for the equation
of state $w_\varphi=1$ corresponds to the \textit{asymptotic in the past} value for the quintessential
exponential potential. Thus, we want to stress again that these behaviours can be naturally connected, and the
exponential form of the scalar field potential driving the late stage of the universe could indeed be
the asymptotic late time behaviour of the inflationary scalar field, which transits from inflation into kination,
toward the late quintessential stage. It is worth noting that such a conclusion is somehow independent of the
mechanism proposed for the evolution of the scalar field potential, being only based on a parametric description
of the very early inflationary dynamics of the universe, and on the property of the equation of state
$w_\varphi=1$, which characterizes our exponential potential. In our cited  forthcoming paper we are also
going to select, among all solutions, the cases which \textit{preserve} such asymptotic behaviour   $w_\varphi=1$.

Of course, this mechanism for driving the transition from the inflationary evolution toward the late time
accelerated expansion has to be considered as mainly exploratory, and some topics need more investigation. For
example, the particles production, that is the gravitational production during the reheating or the preheating,
in which particles are produced by virtue of the variation of the classical inflaton field, needs to be investigated
in a forthcoming paper, considering the scalar field potential given by Eq. (\ref{gamma4}). (As to this, we have
to remind that, in order to study preheating, it is necessary to couple the classical inflaton field $\varphi$
to a massless quantum scalar field $\chi$, as for instance in the Lagrangian
\begin{equation}
{\cal L}_{coupled}= \frac{1}{2} \partial_\mu \chi \partial^\mu \chi - \frac{1}{2} \xi R \chi^2 -\frac{1}{2}
g^2\varphi^2\chi^2\,,
\end{equation}
where $R$ is the scalar curvature, $\xi$ is the gravitational coupling and $g$ is a quartic coupling constant
between $\phi$ and $\chi$.)

To conclude, let us note that considering radiation in the model only changes the evolution of
the $\Omega$ parameters. Radiation initially dominates on matter, while later on the energy densities of matter
and radiation become equal; after that, for some time matter dominates in the universe, while dark energy starts
dominating the expansion rate of the universe only afterwards. These results seem to confirm other
investigations of effects of radiation on the evolution of the quintessence field, according to which the energy
density of the scalar field during the epoch of nucleosynthesis ($z\sim 10^{9}$) is much smaller than the energy
density of radiation. While the kinetic term in the scalar field energy density vanishes and the potential term
becomes constant, the dark--energy term behaves like an effective cosmological constant, not affecting the
process of primordial nucleosynthesis.

\section*{Acknowledgments}
G. Esposito and C. Rubano are grateful to the Dipartimento di Scienze Fisiche of Federico II University, Naples,
for hospitality and support.

\end{document}